\documentclass[letterpaper,useAMS,usenatbib]{mn2e}
\usepackage[pdftex]{graphicx}
\usepackage{amsmath}

\title[GTC OSIRIS transiting exoplanet atmospheric survey]{GTC OSIRIS transiting exoplanet atmospheric survey: 
  detection of sodium in XO-2\lowercase{b} from differential long-slit spectroscopy\thanks{Based on observations made with the Gran Telescopio Canarias
(GTC), installed in the Spanish Observatorio del Roque de los
Muchachos of the Instituto de Astrofísica de Canarias, in the island of
La Palma, and part of the large ESO program 182.C-2018.}}

\author[D. K. Sing]
{D. K. Sing$^{1}$\thanks{E-mail: sing@astro.ex.ac.uk}, 
C. M. Huitson$^{1}$, 
M. Lopez-Morales$^{2}$, 
F. Pont$^{1}$, 
J.-M. D{\'e}sert$^{3}$,\newauthor  
D. Ehrenreich$^{4}$,
P. A. Wilson$^{1}$,
G. E. Ballester$^{5}$, 
J. J. Fortney$^{6}$, \newauthor 
A. Lecavelier des Etangs$^{7}$, and
A. Vidal-Madjar$^{7}$\\
$^{1}$Astrophysics Group, School of Physics, University of Exeter, Stocker Road, Exeter, EX4 4QL\\ 
$^{2}$Institut de Ciencies de l'Espai (CSIC-ICE), Campus UAB, Facultat de Ciencies, Torre C5, parell, 2a pl, E-08193 Bellaterra, Barcelona, Spain\\
$^{3}$Harvard-Smithsonian Center for Astrophysics, Cambridge, MA 02138\\
$^{4}$UJF-Grenoble 1 / CNRS-INSU, Institut de Planétologie et d'Astrophysique de Grenoble (IPAG) UMR 5274, Grenoble, F-38041, France\\
$^{5}$Lunar and Planetary Laboratory, University of Arizona, Sonett Space Science Building, Tucson, AZ 85721-0063, USA\\
$^{6}$Department of Astronomy and Astrophysics, University of California, Santa Cruz, CA 95064, USA\\
$^{7}$Institut d'Astrophysique de Paris, CNRS; Universit\'{e} Pierre et Marie Curie, 98 bis bv Arago, F- 75014 Paris, France\\
}
\begin{document}

\date{Accepted 2012 August 15. Received 2012 July 27; in original form 2012 June 20}

\pagerange{\pageref{firstpage}--\pageref{lastpage}} \pubyear{2012}

\maketitle

\label{firstpage}

\begin{abstract}
We present two transits of the hot-Jupiter exoplanet XO-2b using the Gran Telescopio Canarias (GTC).  The time series observations were performed using long-slit spectroscopy of XO-2 and a nearby reference star with the OSIRIS instrument, enabling differential specrophotometric transit lightcurves capable of measuring the exoplanet's transmission spectrum.  Two optical low-resolution grisms were used to cover the optical wavelength range from 3800 to 9300~\AA.  We find that sub-mmag level slit losses between the target and reference star prevent full optical transmission spectra from being constructed, limiting our analysis to differential absorption depths over $\sim1000$~\AA\ regions.  Wider long slits or multi-object grism spectroscopy with wide masks will likely prove effective in minimising the observed slit-loss trends.
During both transits, we detect significant absorption in the planetary atmosphere of XO-2b using a 50~\AA\ bandpass centred on the Na\,{\sevensize I} doublet, with absorption depths of  
$\Delta (R_{pl}/R_{\star})^2=$ 0.049$\pm$0.017\% using the R500R grism and 0.047$\pm$0.011\% using the R500B grism (combined 5.2-$\sigma$ significance from both transits).  The sodium feature is unresolved in our low-resolution spectra, with detailed modelling also likely ruling out significant line-wing absorption over an $\sim$800~\AA\ region surrounding the doublet.  Combined with narrowband photometric measurements, XO-2b is the first hot Jupiter with evidence for both sodium and potassium present in the planet's atmosphere.  
\end{abstract}


\section{Introduction}

Using transiting exoplanets to obtain transmission spectroscopy is now a proven method to probe planetary atmospheres.  
To date, most of the observed atoms and molecules identified with this technique
have been for the two most favourable cases, the hot Jupiters HD~209458b and HD~189733b.  The detections include 
species in the upper escaping atmosphere such as hydrogen, oxygen, carbon, and silicon 
\citep{2003Natur.422..143V, 
2004ApJ...604L..69V, 
2010A&A...514A..72L, 
2010ApJ...717.1291L, 2012ApJ...751...86J}. 
In the lower atmosphere, sodium has been detected in both of these planets 
\citep{2002ApJ...568..377C, 
2008ApJ...673L..87R, 
2008A&A...487..357S} 
with the line profile also able to place constrains on the temperature-pressure profile 
\citep{2008ApJ...686..667S, 
 2011A&A...527A.110V, 
vidalmadjar11b,
2012MNRAS.422.2477H}.  
Alkali metals in other hot Jupiters have been detected as well, with evidence for sodium in Wasp-17b
\citep{2011MNRAS.412.2376W} 
and potassium in XO-2b 
\citep{2011A&A...527A..73S}. 

Contrary to clear hot-Jupiter atmospheric models, HD~189733b shows ample evidence for a dominant haze which extends from the optical 
\citep{2008MNRAS.385..109P, 
2011MNRAS.416.1443S} 
into the near-IR 
\citep{2009A&A...505..891S, 
2009ApJ...699..478D, 2011A&A...526A..12D, 
2012MNRAS.422..753G}. 
The signature of the haze is characteristic of Rayleigh scattering by small particles 
\citep{2008A&A...481L..83L}, 
with silicate grains a possibility as predicted by some cloud models 
\citep{
1999ApJ...513..879M, 
2000ApJ...538..885S, 
2000ApJ...540..504S, 
2008A&A...485..547H}. 

In the context of hot Jupiters, the alkali metals sodium and potassium are extremely important.  Alkali metals are 
thought to be responsible for the low albedos of hot Jupiters, with the pressure broadened potassium and sodium 
line-wings efficiently absorbing the incoming optical stellar radiation 
\citep{2000ApJ...538..885S, 2000ApJ...540..504S}.  
Most hot Jupiters do appear to have low albedos \citep{2011ApJ...729...54C}, 
though there are exceptions 
\citep{2011ApJ...735L..12D, 2011A&A...536A..70S}. 
As the alkali metals are thought to play such a dominant role in the planetary energy budget of hot Jupiters, dictating where the 
stellar radiation is deposited, placing observational constraints on these metals is an important first step.

Sodium and potassium can both be detected from space-based or ground-based spectra or photometric measurements (e.g. \citealt{2002ApJ...568..377C,2008ApJ...673L..87R,2011A&A...527A..73S}).
The challenge of obtaining broadband transmission spectroscopy from the ground is the requirement for very high photometric accuracy, along with good spectral resolution on both the target and reference stars.  Observations toward this end have been made on large aperture telescopes with integral field spectroscopy
\citep{2006PASP..118...21A} 
and multi-object spectrographs
\citep{2010Natur.468..669B, 2011ApJ...743...92B}. 
Here, we report the use of long-slit time-series spectroscopy to take simultaneous measurements of a transit event along with a nearby reference star, which is used for differential spectrophotometry.  Such multi-object spectroscopic techniques have the potential to deliver transmission, emission, and albedo spectra of transiting extrasolar planets, making such observations competitive with space-based instruments.  The larger telescope apertures available from the ground also lead to better time sampling of the short hour-long transit events and the ability to observe fainter targets, an important step in being able to characterise the large number of fainter transiting planets discovered from photometric surveys.  The observations reported here are part of a large ESO programme on the GTC, performing a spectrophotometric optical
survey of transiting hot Jupiters using narrowband imaging and long-slit spectroscopy.  
In this paper, we describe the GTC XO-2b observations in \S 2, present the analysis of the long-slit transit
light curves in \S 3, discuss the results and compare to model atmospheres in \S 4, and conclude in \S 5.

\section{Observations}
We observed the exoplanet host star XO-2A with the GTC 10.4 metre telescope installed in the Spanish Observatorio 
del Roque de los Muchachos of the Instituto de Astrof\'\i sica de Canarias on the island of La Palma, 
using the Optical System for Imaging and low Resolution Integrated Spectroscopy (OSIRIS) 
instrument during two separate nights in service mode.

\subsection{GTC OSIRIS instrument setup}
The GTC OSIRIS instrument \citep{
2000SPIE.4008..623C, 
2003SPIE.4841.1739C} 
consists of two Marconi CCD detectors, 
each with 2048$\times$4096 pixels and a total
FOV of 7.8$'\times$8.5$'$, giving a plate scale of 0.127$''$/pixel.  The
CCD has a pixel well depth of $\sim$100,000 electrons and a 16 bit A/D
converter, which saturates at 65,536 counts.  For our observations, we
choose the 2$\times$2 binning mode with a full frame readout at a speed of
500kHz, which produces a readout overhead time of 17 seconds.
The readout speed of 500kHz has a gain of 1.46
e$^{-}$/ADU and a readout noise of 8 e$^{-}$.  While 500kHz is the 
noisiest readout speed (100 and 200kHz are available), it allows for
the most counts per image to be obtained due to a higher gain and a shorter readout time which provides a better sampling rate.  

We utilised the spectroscopic capabilities of OSIRIS, using a long slit of 8.67$'$ in the spatial direction with a 5$''$ width.  We used the widest available slit at the time of the observations to minimise slit losses.  We observed two transit events of XO-2b, observing one with the R500B and one with the R500R grism.  The R500B grism covered a spectral range from 3750 to 8586 \AA, has a dispersion of about 3.7~\AA\ per binned pixel, and has a peak efficiency of 17\% near 5000\AA.  The R500R grism covered a spectral range from 5000 to 9300 \AA, has a dispersion of about 4.7 \AA\ per binned pixel, and has a peak efficiency of 15\% near 6000\AA.  

\begin{figure}
 {\centering
  \includegraphics[width=0.49\textwidth,angle=0]{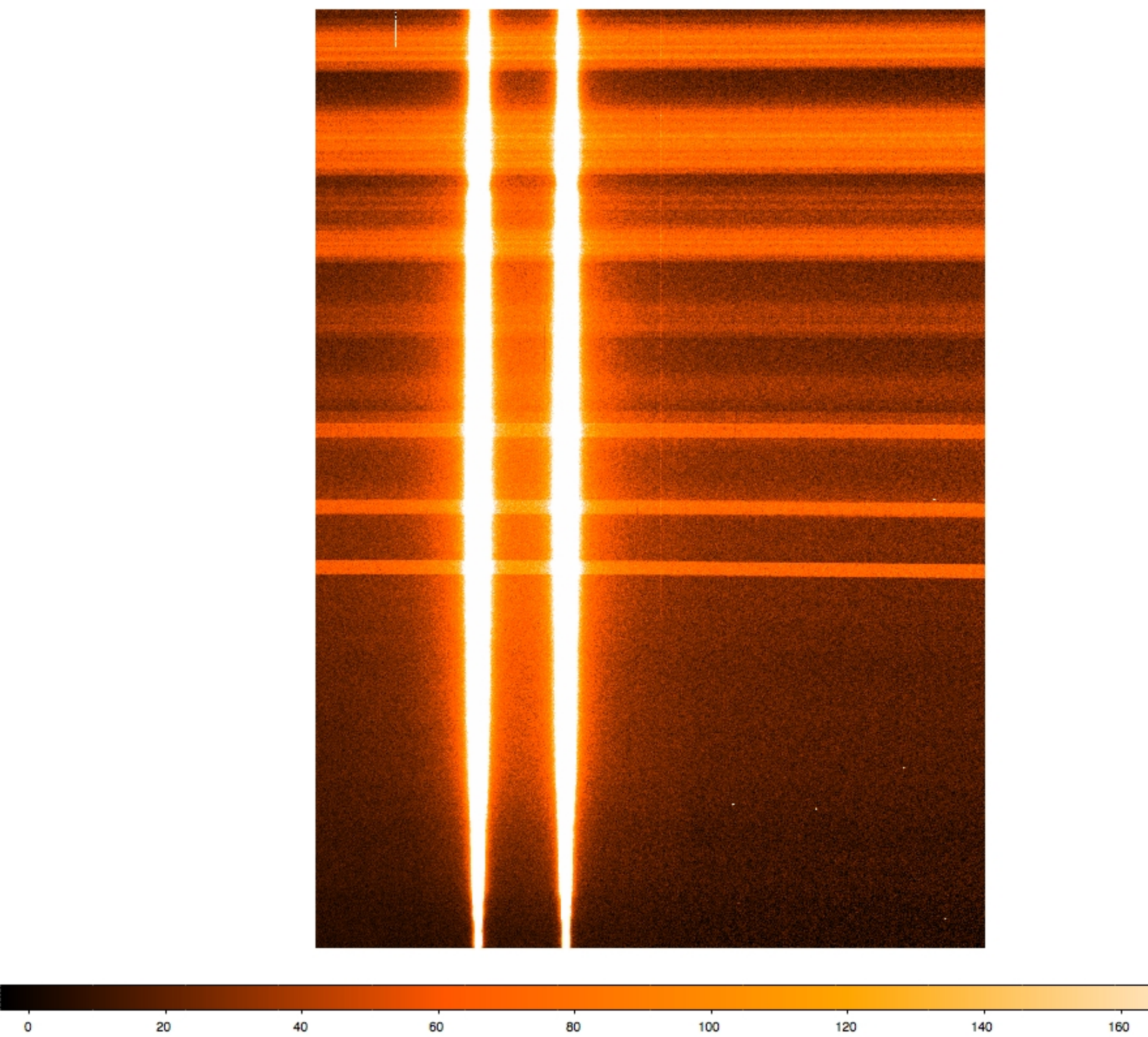}}
\caption[]{GTC Osiris CCD frame of XO-2A (right) and XO-2B (left) long-slit spectra from 23 Feb 2012 taken with the
  R500B grism and a 5'' wide slit.  The spectra cover wavelength ranges from 3750~\AA\ (bottom) to 8587~\AA\ (top), with background sky emission lines seen in the cross-dispersion direction.}
\label{Figure:FOV}
\end{figure}

For the XO-2b observations, we placed both the target and a chosen nearby reference star, XO-2A's stellar companion XO-2B, within the slit, orientating the spatial direction such that both stars were centred within the 5$''$ slit (see Fig. \ref{Figure:FOV}).  The companions XO-2A and XO-2B are separated by only 31$''$ and cataloged as each having the same spectral type (K0V) and have very similar magnitudes and colours, making XO-2B an ideal reference star.  In the future, multi-object spectroscopy modes for OSIRIS are planned, which will provide an added benefit over long-slit spectroscopy as more reference stars will be obtainable and wider slit masks can be fabricated.  In addition, wider long slits are planned for OSIRIS as well.

\subsection{Observing log}
XO-2A and XO-2B were observed keeping them simultaneously on the slit during the XO-2b transit events of 23 December 2011 and 23 February 2012.  
 
{\it 23 Dec. 2011}:  The seeing was stable during the observations, with values ranging from 0.8$''$ to 0.9$''$ under photometric conditions.  
The R500R grism was used with 10 second exposure times, with 770 exposures obtained between UT 01:01 and UT 06:56, with airmass values ranging between 1.07 and 1.54.  

{\it 23 Feb. 2012}:  During the night, the seeing ranged from 0.8$''$ to 2.0$''$ under photometric conditions.  
The R500B grism was used with 22 second exposure times, with 519 exposures obtained between UT 20:32 and UT 02:15, with airmass values ranging between 1.07 and 1.4.  

For the two nights, exposure times were not modified during the sequence, maintaining the peak counts between 25000-40000 ADUs for both the reference and target stars, well within the linear regime of the detectors.  When needed, defocusing was used to avoid saturation in improved seeing conditions, with the full width at half maximum (FWHM) of the point spread function (PSF) typically kept about $\sim$2.5 times smaller than the slit width.  
At the expense of higher spectral resolution, adopting a modest defocus also helps to minimise changes to the spectral resolution.
For the 23 Feb. observations, 97\% of the spectra have a FWHM within 25\% of the average, despite a factor of more than two variation in the reported seeing conditions.  The wavelength calibration was determined from HgAr, Xe and Ne arc lamps.
For each night 100 bias and 100 well exposed dome flat-field images were taken ($\sim$30,000 ADU/pixel/image).  Our program uses
dome-flats, as opposed to sky-flats, as small scale pixel-to-pixel variations are a potentially important source of noise, requiring
a large number of well exposed flats to effectively remove it.  Our
observations were all specified to occur at the same pixel location, to help suppress flat-fielding errors.  Through ESO GTC program 182.C-2018, we used 12 transit light curves taken with narrowband imaging to estimate the guiding performance during a typical 4.5 hour long transit observation. We found the
guiding performance of GTC OSIRIS to be quite stable, with measured drifts of 1.4$\pm$1.0 pixels (0.18$\pm0.13$$''$)
in the CCD X-direction and 1.0$\pm$0.6 pixels (0.13$\pm0.06$$''$) in the Y-direction over a 4 hour period.

\subsection{Reduction}

\begin{figure}
 {\centering                                                                      
  \includegraphics[width=0.4\textwidth,angle=90]{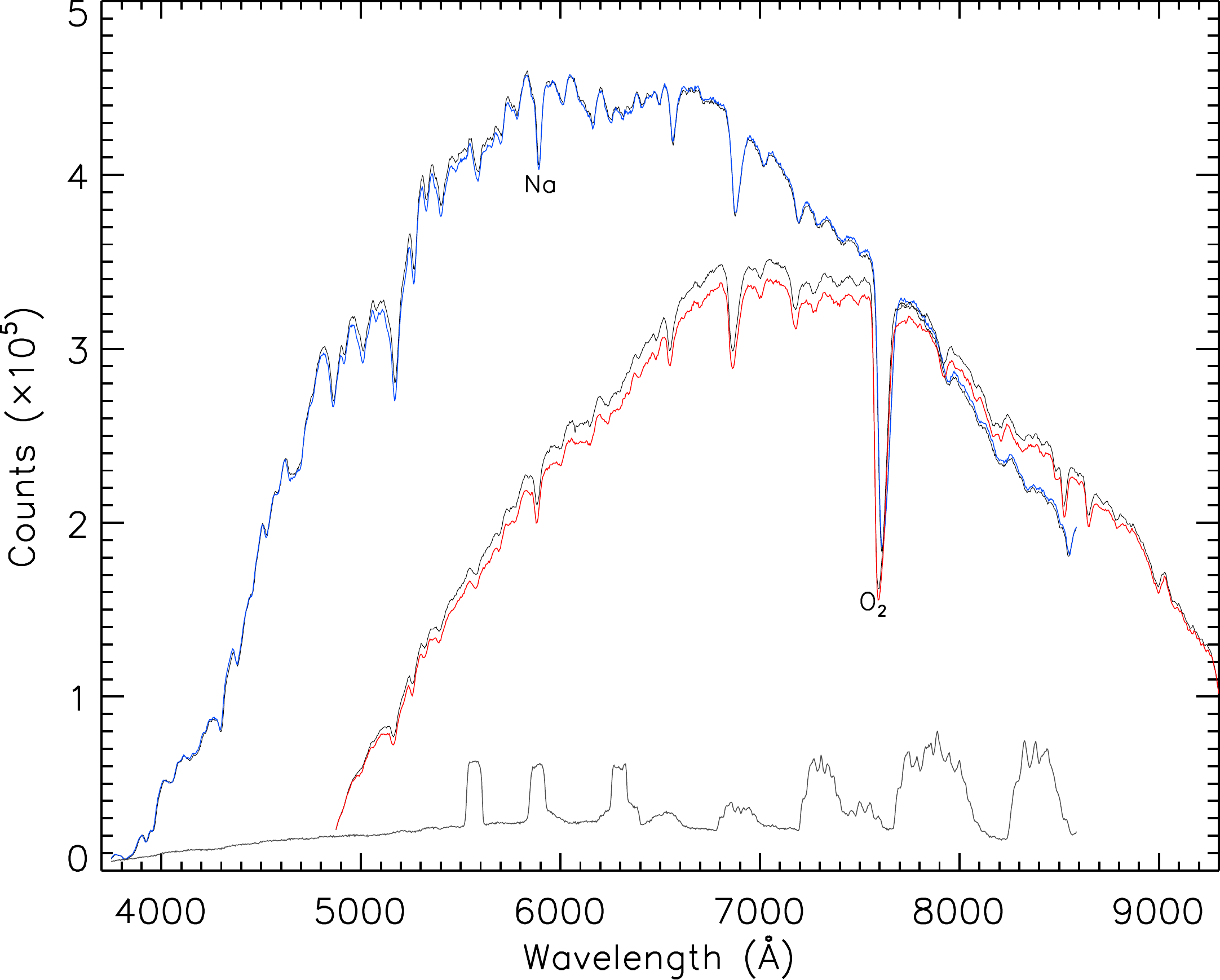}}
\caption[]{Extracted R500B and R500R grism spectra of XO-2A (blue \& red) and the reference star XO-2B (black).  Also overplotted (grey) is a R500B spectra of the sky background, increased in counts by a factor of 10 for illustration purposes.}
\label{Figure:spec}
\end{figure}

The bias frames and flat fields were combined using standard IRAF routines, and used to correct each image.  
Aperture extraction for both the target and reference spectra was done using a wide 
$\sim$40 pixel aperture and background subtraction (see Fig. \ref{Figure:spec}).  For both nights, the appropriate aperture size was chosen to minimise the out-of-transit light curve dispersion.
The extracted spectra were then used to create differential spectrophotometric light curves, summing the spectra over wavelength regions of interest, and dividing the flux of the target star by that of the reference.  With a low-resolution grism in conjunction with a wide slit and modest defocus, the spectra are inherently of low spectral resolution limiting bandwidths to $\sim$30 \AA\ or larger (R=200), with 50\AA\ bands the narrowest chosen for this study (R=120).  

\section{Analysis}
We modelled the transit light curves with the analytical transit
models of \cite{
2002ApJ...580L.171M},  
fitting the integrated light curves for the central transit time, planet-to-star radius contrast, inclination, stellar density, and baseline flux using a Levenberg-Marquardt least-squares algorithm \citep{2009ASPC..411..251M}. 
To account for the effects of limb-darkening on the transit light curve, we adopted the three parameter limb-darkening law, 
\begin{equation}  \frac{I(\mu)}{I(1)}=1 - c_2(1 - \mu) - c_3(1 - \mu^{3/2}) - c_4(1 - \mu^{2}) \end{equation}
calculating the coefficients following \cite{
2010A&A...510A..21S}. 
The errors on each datapoint were set to the standard deviation of the fit to the out-of-transit residuals. 
Following \cite{
2006MNRAS.373..231P}, 
we assessed the levels of residual red noise by modelling the binned variance with a $\sigma^2=\sigma_w^2/N+\sigma_r^2$ relation, 
where $\sigma_w$ is the uncorrelated white noise component, $N$ is the number points in the bin, and $\sigma_r$ characterises the red noise.  
We subsequently rescaled the fitted parameter uncertainties to incorporate the measured values of $\sigma_r$ in the white-light and spectral bin fits.  

In our fits, we determined the best-fit system parameters (inclination, central transit time, ect.) from the white-light curve before measuring the differential radius variations of the transmission spectra using wavelength bins.

\subsection{Broadband transit light curves and slit losses}
\begin{figure}
 {\centering                                                                      
  \includegraphics[width=0.35\textwidth,angle=90]{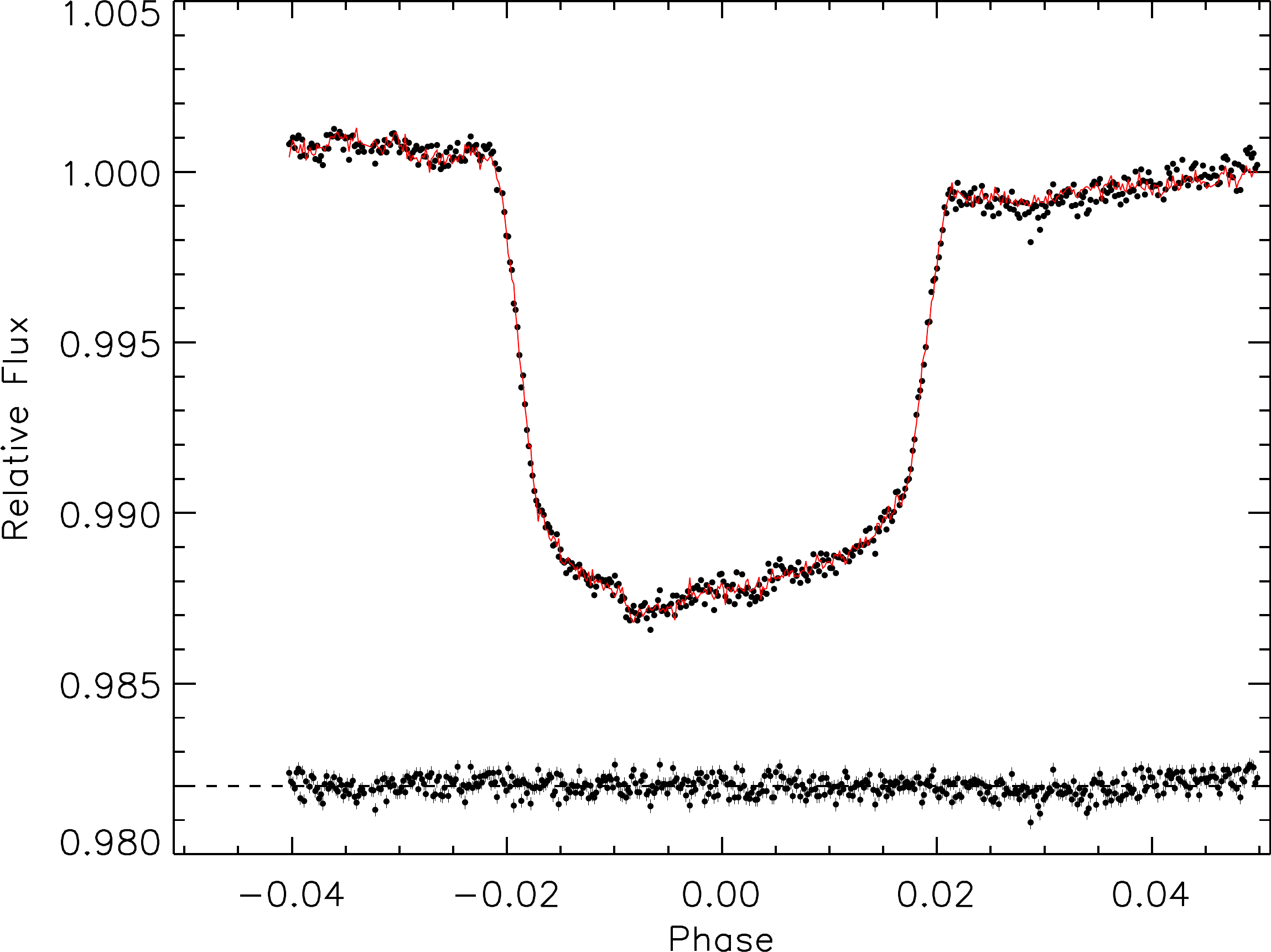}}
\caption[]{The whitelight transit light curve using the R500B grism (black).  Also overplotted is the fitted model light curve including parameterised systematic trends (red) and the residuals with error bars (black points bottom).}
\label{Figure:white}
\end{figure}

We summed up the R500B spectrum from 4066 to 7590 \AA\ to characterise the Feb. transit white-light curve (see Fig. \ref{Figure:white}).   Summed over a wide spectral area, each of the white-light curve points has a very high total count rate, typically $3 \times 10^8$ counts per image, which gives a theoretical photon-limited precision of 7$\times 10^{-5}$.    The observed transit light curve, however, displays clear mmag-level systematic trends which vary on a $\sim$2.5 hour timescale.  
We attribute these trends to differential slit losses between the target and reference star.  

\begin{figure}
 {\centering                                                                      
  \includegraphics[width=0.35\textwidth,angle=90]{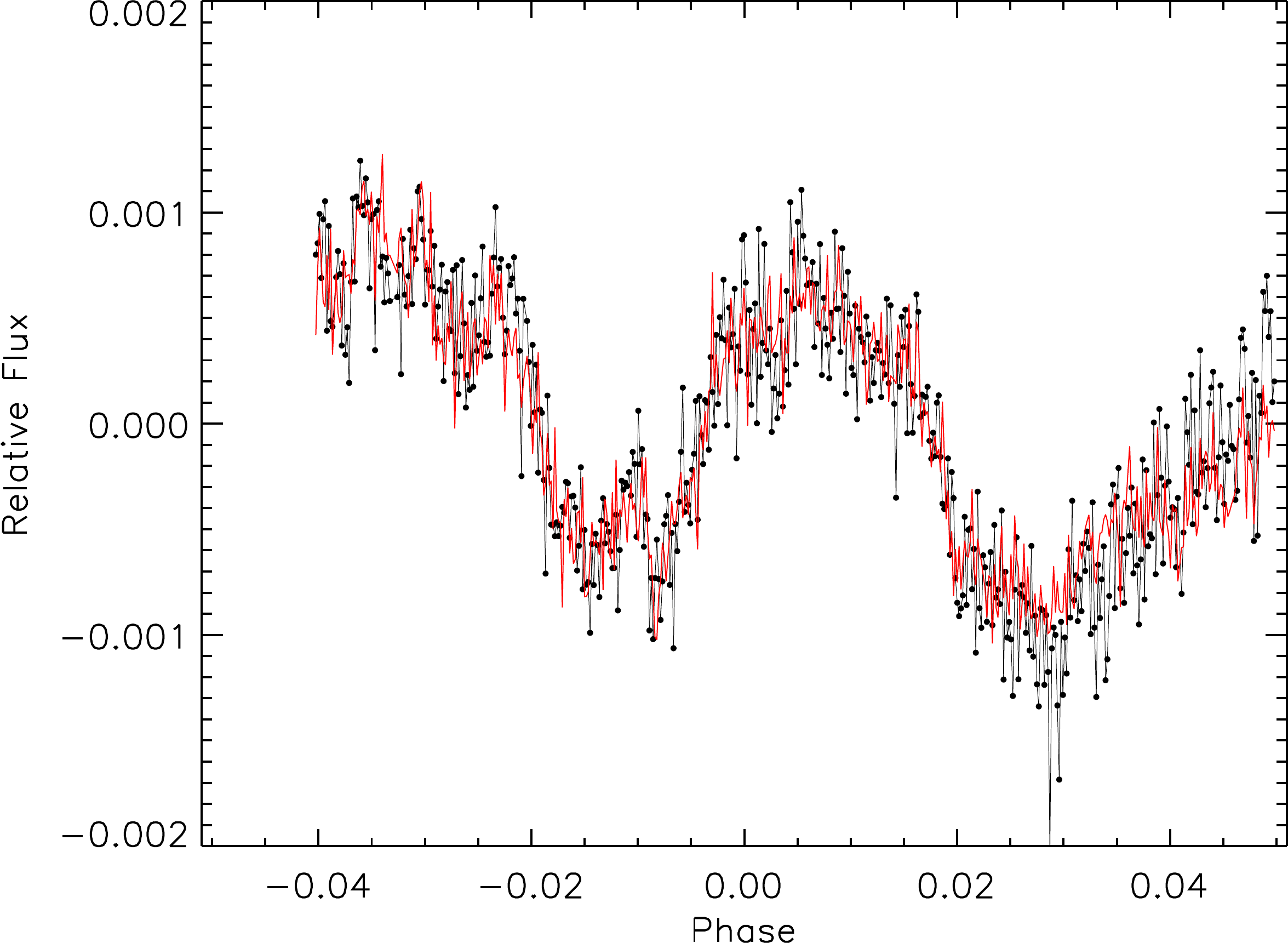}}
\caption[]{The Feb. R500B light curve with the transit removed illustrating the slit-losses (black) also overplotted with a parameterised model (red).}
\label{Figure:slitlosses}
\end{figure}

The systematic slit-loss trends of the Feb. transit light curve exhibit a regular sinusoidal trend (see Fig. \ref{Figure:slitlosses}).  In discussions with the GTC staff, the probable cause of the regular and repetitive oscillation observed is the telescope guiding (A. Cabrera-Lavers priv. comm.).  As the two objects are orientated following a specific slit-position angle and the guiding is performed with respect to the target, there are subtle slit losses due to the lack of rotator correction in the guiding.  The guiding corrects by way of a linear displacement in the focal plane, and as no additional rotation is done, the effect in the non-target star will be translated as a small positional displacement that can be approximated as sine and cosine functions in the focal plane.   With this interpretation, the regular trends in Fig. \ref{Figure:slitlosses} are the result of differences in slit losses between the target and reference star, primarily due to guiding errors.

\subsection{Parameterised slit losses}
To model the systematic trends of the Feb. R500B light curve, we parameterised the slit losses with a combination of a sine and cosine function as a function of time, fitting for the sinusoidal amplitudes $a$, periods $p$, and phases $\theta$.  Upon searching for trends in other external parameters, such as CCD position and sky brightness, a linear full width at half maximum (FWHM, $\delta_w$) trend and linear base trend in orbital phase, $\phi$, were added as well.  Shown in Fig. \ref{Figure:slitlosses}, the total time-FWHM-dependent baseline flux $f(\phi,\delta_w)$ was described by a fit with nine free parameters, eight describing the systematic trends and one measuring the differential system flux $f_0$,
\begin{multline}
f(\phi,\delta_w)=f_0 \times (c_1\phi+1) \times
(a_1\sin(2\pi \phi / p_1+\theta_1)+1)\times\\
(a_2\cos(2\pi \phi / p_2+\theta_2)+1) \times (c_2\delta_w+1)
\end{multline}
As with the sinusoidal trends, a FWHM dependance is also expected for slit losses, as wider Point Spread Functions (PSF) lead to wider wings and larger slit losses.

The cosine term captured a higher harmonic of the overall 2.5 hour modulation, as the fitted parameters were close to half the period, half the amplitude, and 90 degrees out of phase compared to the fitted sinusoidal terms.  Further higher order (higher frequency) harmonics are evident in the residuals, which have an amplitude of $\sim$2$\times 10^{-4}$ with periods all shorter than $\sim$13 minutes.  The sine and cosine terms effectively remove the hour-long timescale trends, relevant for measuring the planetary radius. 

With the parameterised slit losses, the standard deviation of the residuals is estimated to be 2.3$\times 10^{-4}$ for the R500B white-light curve, which is dominated by the residual slit-loss trends and a factor of three above the theoretical photon noise limit.  
We find a system inclination of 88.8$\pm$0.9 deg and stellar density of 1.5$\pm$0.1 g cm$^{-1}$ for the R500B transit (see Table \ref{Table:bestfit}).  These values match well with previous determinations, including the GTC photometric measurements of \cite{2011A&A...527A..73S} who found 87.62$\pm$0.51 deg and 1.328$\pm$0.088 g cm$^{-1}$.

To measure the transmission spectra, we also sub-divided the R500B spectra into different sub-bins of $\sim$50~\AA\ and wider.  Parameterising the slit losses comes close to providing a reliable transmission spectrum, which would be preferable as absolute transit depths would be measurable.  However, we find that residual slit losses are still present at an unacceptably high level.  The relevant radius precision one needs to observe atmospheric features is about one atmospheric scale height, estimated to be $H$=370 km \citep{2011A&A...527A..73S}.  This target gives a required planet-to-star radius contrast precision of $H/R_{\star}$=0.00054 which is an absorption depth of $\Delta (R_{pl}/R_{\star})^2$=0.01\%.  With the adopted parameterisation, the residual slit-loss trends and limiting precision is about twice this target value.  Further observations with wider slits will reduce these observed slit-loss trends.  With GTC, based on the observed pointing accuracy and PSF in good seeing conditions, 10$''$ slits or wider should encompass the entire PSF and lead to minimal slit losses.

\begin{figure}
 {\centering                                                                      
  \includegraphics[width=0.35\textwidth,angle=90]{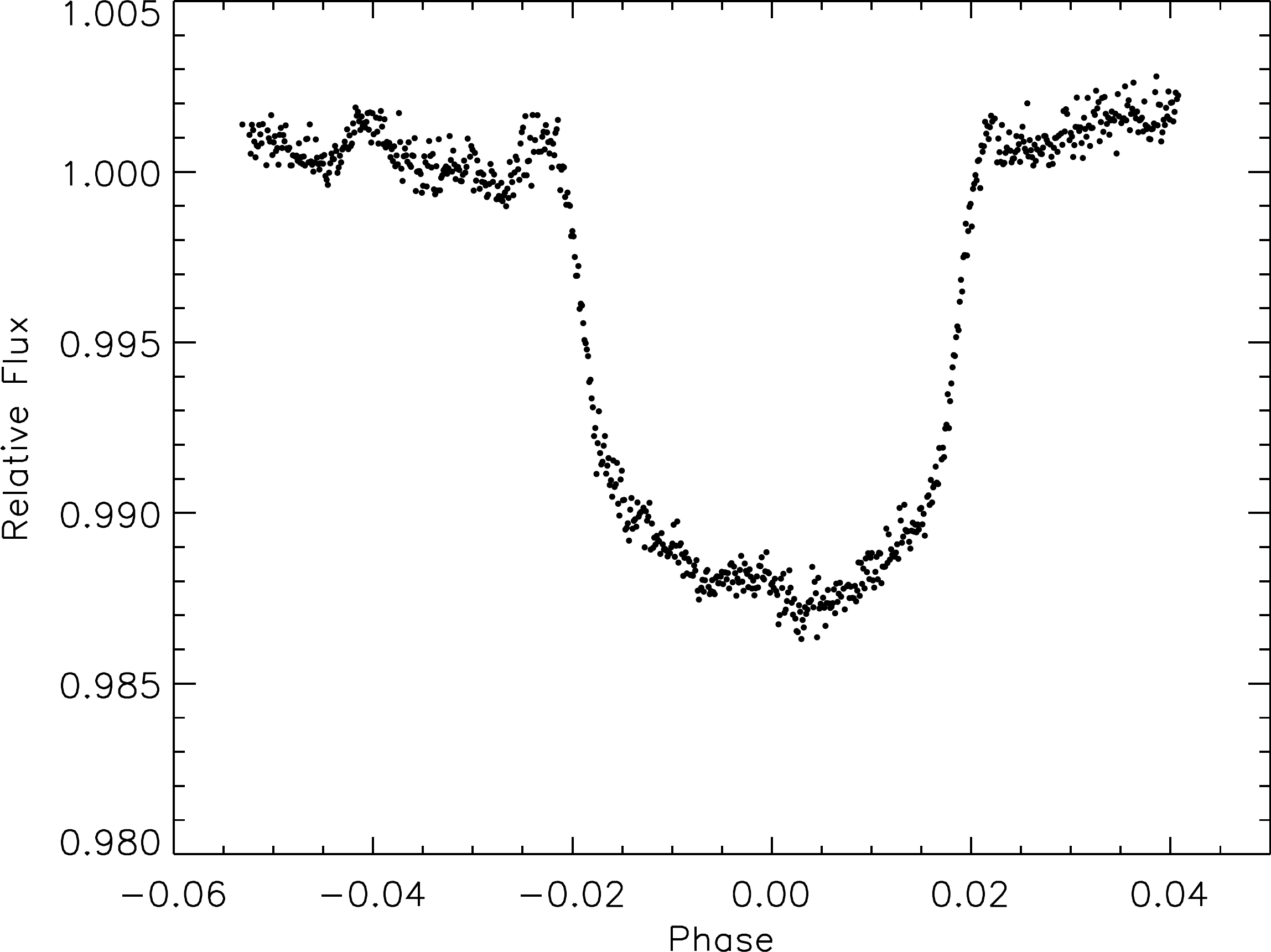}}
\caption[]{The whitelight transit light curve using the R500R grism.}
\label{Figure:whiteR}
\end{figure}

The R500R observation (see Fig. \ref{Figure:whiteR}) was best fit with a baseline flux and a linear trend with phase, a linear trend with X-position, plus a quadratic trend with FWHM.  However, the fits are of poor quality as the slit-loss trends are of much higher frequency, and are difficult to fit without including a large number of modes.

\begin{table} 
\caption{Best fit parameters for the XO-2b R500B transit}
\label{Table:bestfit}
\begin{centering}
\renewcommand{\footnoterule}{}  
\begin{tabular}{lll}
\hline\hline  
Parameter & Value \\
\hline 
Period, $P^\dagger$ [days] &2.61586178$\pm$0.00000075 \\
Mid-transit Time, [BJD]  & 2455981.46036$\pm$0.00013 \\
radius, R$_{pl}$/R$_{\star}$        &   0.1052$\pm$0.0011   \\
inclination, $i$ [deg]                                               &    88.8$\pm$0.9   \\
system scale, $a$/R$_{\star}$                                   &   8.17$\pm$0.08  \\
impact parameter, $b$=$a$cos$i$/R$_{\star}$        &    0.174$\pm$0.060   \\
stellar density, $\rho_{\star}$ [g cm$^{-3}$]              &   1.507$\pm$0.1   \\
limb-darkening coeff., $c_{2}$                      &  ~1.3036  \\
limb-darkening coeff., $c_{3}$                      &  -0.8579\\
limb-darkening coeff., $c_{4}$                      &  ~0.3005 \\
\hline
\multicolumn{2}{c}{Previously determined model dependent parameters}\\
\hline
$R_{\star}$ [$\mathrm{R_{\odot}}$]    &  0.976$\pm$0.020$^{\ddag}$ \\
$R_{pl}$  [$\mathrm{R_{Jup}}$]  & 0.996$\pm$0.025$^{\ddag}$ \\
$M_{\star}$ [$\mathrm{M_{\odot}}$]    &  0.971$\pm$0.034$^{\ddag}$ \\
$M_{pl}$ [$\mathrm{M_{Jup}}$]   &  0.565$\pm$0.054$^{\ddag}$ \\
\hline
\end{tabular}
\end{centering}
\\
$^\dagger$from Sing et al. (2011), fit along with the data from \cite{2007ApJ...671.2115B}  and \cite{2009AJ....137.4911F}\\  
$^{\ddag}$From \cite{2009AJ....137.4911F} \\
\end{table}

\subsection{Common-mode slit losses and differential transmission spectra}
As both the R500R and R500B lightcurves show multiple higher-order modes, requiring a prohibitory large number of free parameters, we also employed a common-mode technique to remove the slit-loss trends.  This technique relies on the similarities of the slit-losses across a spectrum, which can be characterised by the light curves themselves and removed individually for each spectral wavelength bin.  For both the R500B and R500R, we find the overall systematics of the white-light curve dominate over wavelength-dependent trends by a factor of about six.  Subtracting common-mode systematics are particularly useful when measuring transmission spectra, as the relative wavelength-dependent transit depths are usually not effected.  

Empirically determining and removing the slit losses has an advantage over the parameterised method, as the higher order frequencies are naturally subtracted.  However, absolute transit depths cannot be be obtained, and the slit losses do appear to have a slight wavelength dependence brought about through different flux illumination levels.  Higher count rates, at a particular wavelength, result in higher counts in the tails of the PSF, which spill over the 5$''$ slit and lead to larger slit losses.  

\begin{figure}
 {\centering                                                                      
  \includegraphics[width=0.35\textwidth,angle=90]{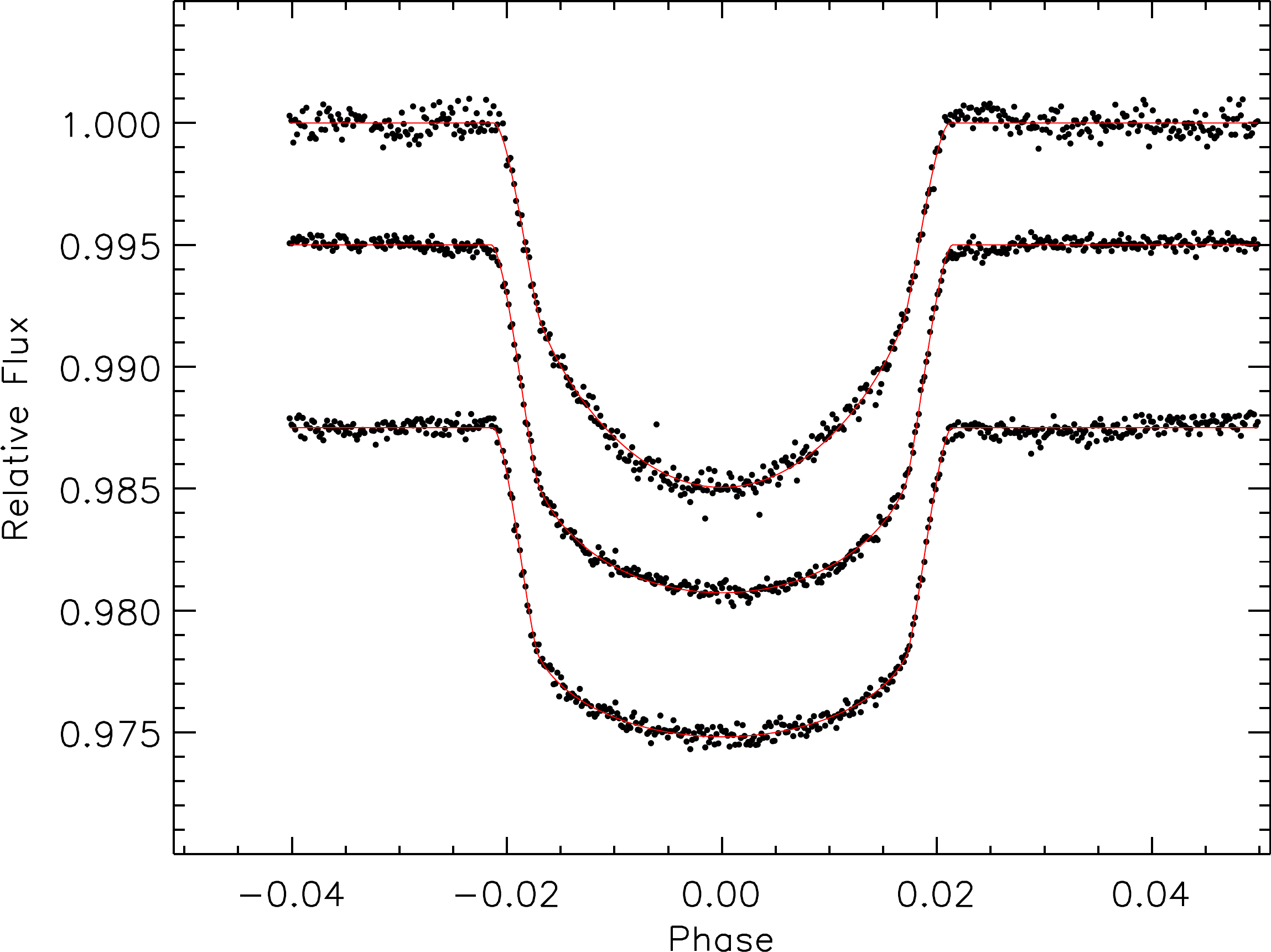}}
\caption[]{Three broadband transit light curves extracted from the R500B grism (black), correcting for slit losses using those from the white-light curve (as seen in Fig. \ref{Figure:slitlosses}), with an arbitrary offset for clarity.  The effects of limb darkening are easily distinguished from the bluest spectral bin (top, 4066 to 4745~\AA ) compared to mid-spectral wavelengths (middle, 5849 to 6612~\AA ) and the reddest bin (bottom, 7790 to 8582~\AA ).  The light curves have precisions of 400, 260, and 210 ppm with a cadence of 39 sec. (top to bottom respectively).}
\label{Figure:broadband}
\end{figure}

We employed different techniques to remove the common-mode trends, including using the slit losses from the white-light curve (e.g. Fig. \ref{Figure:slitlosses}) as an auxiliary decorrelation parameter in the transit light curve fits (see Fig. \ref{Figure:broadband}).  We also used a differential-transit method, subtracting a reference-bin light curve from those of surrounding spectral bins, and fitting for the differential planetary radii $\Delta R_{pl}/R_{star}$.  In this method, the majority of the transit itself is subtracted away, apart from the differences in limb-darkening and planetary radius, but the slit losses themselves are also largely removed.  
Both methods produced similar results, though we quote the results and draw conclusions from the differential-transit fits, as we were able to fit the light curves with a minimum number of free parameters and avoided fitting for nuisance parameters describing the slit losses.  The amplitude of the residual wavelength-dependent slit losses in the differential-transit fits were then estimated from $\sigma_r$ for each wavelength bin, with the errors on the transmission spectra adjusted accordingly. 

The slit-loss dominated trends (e.g. Fig. \ref{Figure:slitlosses}) for the R500B and R500R grisms have standard deviations of 62$\times 10^{-5}$ and 68$\times 10^{-5}$ respectively, and thus have an amplitude which is on the order of the size of the atmospheric spectral features we are searching for.
The common-mode differential-transit technique, which removes a large portion of the slit losses, improves the limiting precision by about a factor of six, with typical values of the residual correlated noise measured to be $\sigma_r \approx 10\times 10^{-5}$.  
The achieved precision is about a factor of two better than the parameterised method, and sufficient to confidently measure and identify atmospheric features.

We quantitatively assessed the spectral extent to which we could use the slit losses of a given reference bin by measuring the linear Pearson correlation coefficient between the out-of-transit portions of the reference and spectral-bin light curves.  Well correlated slit losses produced correlation values of 0.8 and higher, giving red noise values near $\sigma_r=10\times10^{-5}$ in the differential-transit light curves.  We found that spectral ranges were limited to no larger than $\sim$500~\AA\ away from the centre of any particular reference bin, as correlation values drop significantly and the red noise increases.  In the context of hot-Jupiter transmission spectra, this limited our analysis to searches of expected specific atomic features including sodium, potassium, and H$_{\rm\alpha}$.  In our searches, potassium was unfortunately compromised by the large and variable telluric O$_2$ lines (also see \citealt{2012ApJ...746...46C}), and the H$_{\rm\alpha}$ line did not show significant absorption.  

\begin{figure}
\centering
\includegraphics[width=8cm]{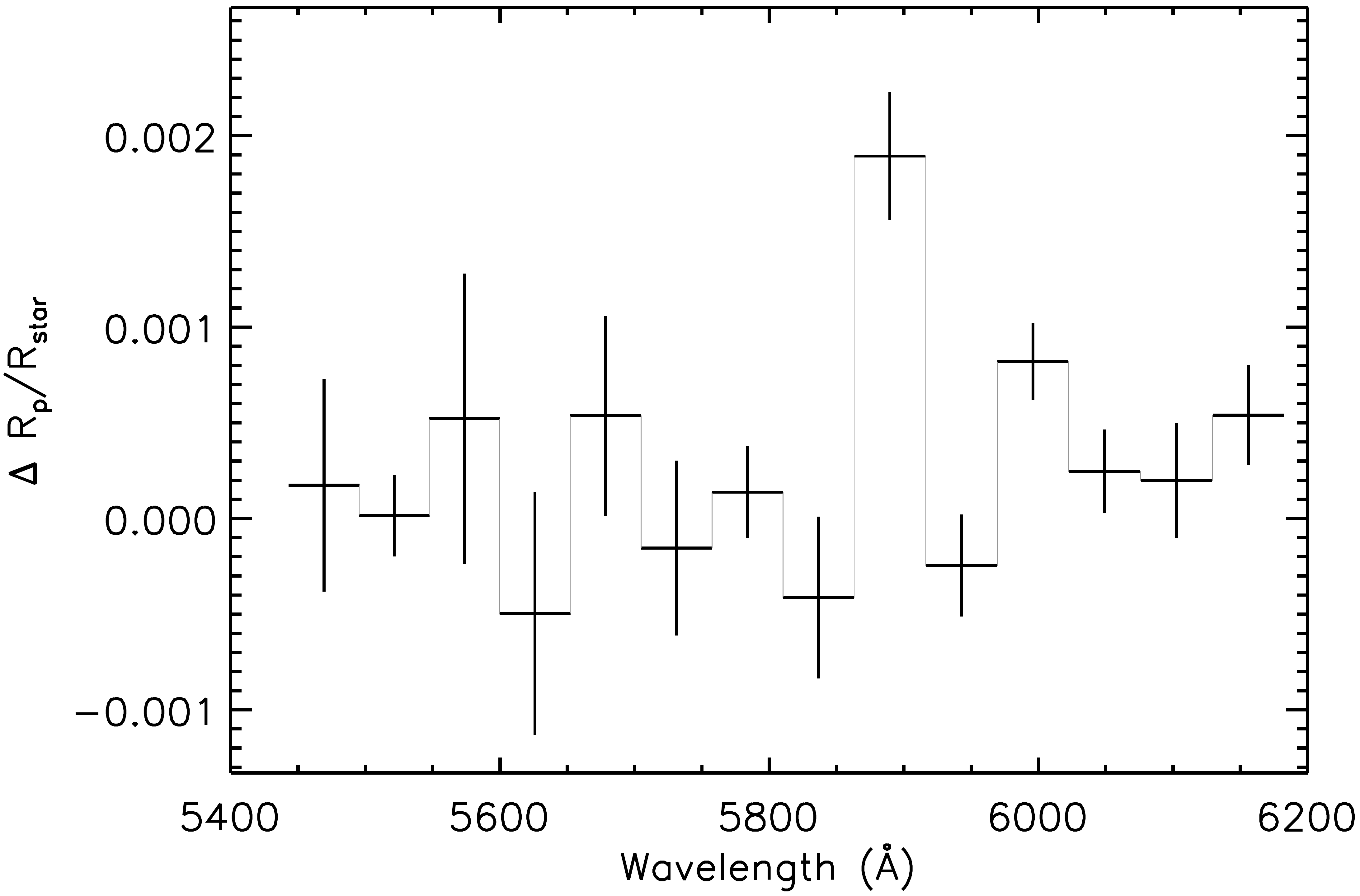}
\caption{The differential spectrum around the sodium feature, using a weighted mean of both observations where the wavelengths overlap (5850-6150 \AA).  The X-axis error bars indicate the wavelength spectral bins, while the Y-axis error bars indicate the 1-$\sigma$ uncertainty in differential radius.} 
\label{differential_spectrum}
\end{figure}

In the sodium search, we restricted our extracted spectrum to the small region around the sodium feature.  The wavelength ranges were 5450-6150~\AA $ $ for the R500B spectrum and 5850-6150~\AA $ $ for the R500R spectrum. The efficiency of the CCD introduces significant errors in the R500R spectrum blue-ward of 5800~\AA, where the efficiency is less than half of that at the peak response.  The light curves were fit adopting the stellar model limb darkening values as well as fixing the ephemeris, inclination, and stellar density to the best-fit values found previously.  The fit included three free parameters, one parameter for the differential planet to star radius $\Delta R_{pl}/R_{star}$ and two parameters for the baseline flux, allowed to vary in time linearly.  The two transits were fit individually, with both transits showing significant absorption in the sodium band.  The resulting combined spectrum is shown in Figure \ref{differential_spectrum}, using bin sizes of 50~\AA. 

We tried excluding the reference star XO-2B from the differential-transit analysis, though found significantly worse results, indicating the reference star does help reduce the impact of slit losses in the photometric light curves.
We also investigated the effect of changing the spectral binning on the resulting differential spectrum in increments of 12~\AA.  Including the variance of each point from this method had a negligible effect on the spectral shape or error bars, except in the sodium feature, where the absorption depth decreased if the band was moved more than 25~\AA $ $ away from the central wavelength of the feature, consistent with a positive detection.  In addition, wider bandwidths dilute the measured sodium absorption depth, with the feature confined to bandwidths narrower than $\sim$50 \AA.  
The red noise was seen vary by up to $\sim 20$~per~cent depending on the particular bins used, and we selected the binning which resulted in the lowest $\sigma_r$.  We also found that changing the reference wavelengths by 50-100~\AA\ made no difference to the shape of the relative spectrum.  
The measured values for $\sigma_r$ derived from the binning technique also correspond well with the observed dispersion of the transmission spectra, excluding the sodium line, giving further indications that the values are accurate.

\begin{figure}
 {\centering                                                                      
  \includegraphics[width=0.35\textwidth,angle=90]{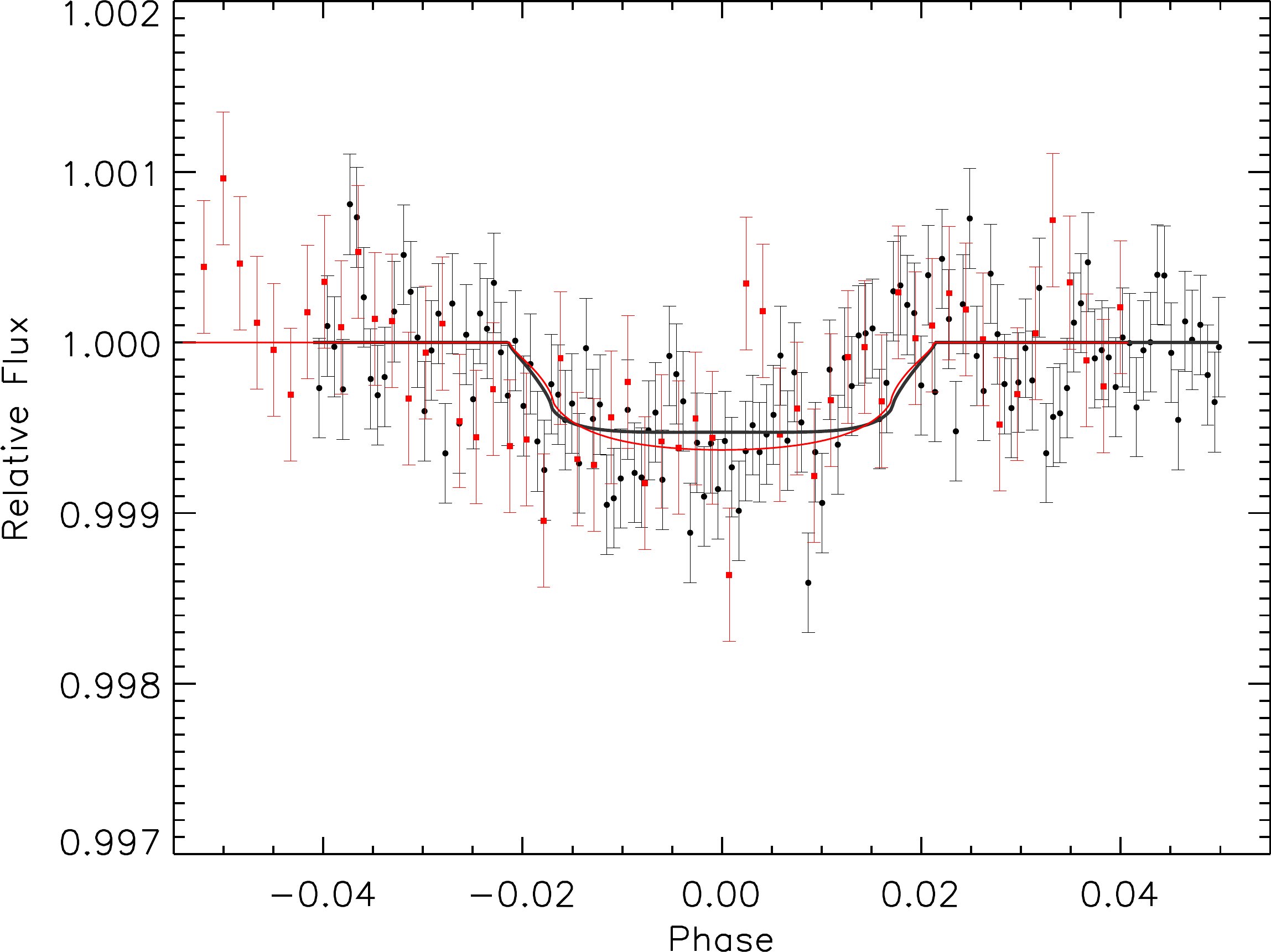}}
\caption[]{Differential transit light curves of a 50 \AA $ $ band centred on the sodium doublet to that of a reference composed of a blue (5812-5862 \AA) and a red (5917-5967 \AA) band which bracket the sodium doublet.  Both transits using the R500B (black circles) and R500R (red squares) are plotted, binned by 4 and 14 points respectively along with the best-fit models (black and red respectively).  The R500R light curve has stronger differential limb-darkening (producing a deeper, more rounded transit) due to the weaker spectral responce in the blue reference band.}
\label{Figure:Nafig}
\end{figure}

We also measured the differential absorption depth of just the sodium feature itself using the differential-transit method, which is similar to the procedures in \citet{2002ApJ...568..377C}.  We measured the differential transit depth between a 50 \AA $ $ band centred on the sodium doublet to a reference composed of a blue (5812-5862 \AA) and a red (5917-5967 \AA) band which bracket the sodium doublet.  The differential lightcurves were first normalised to the out-of-transit flux for the given wavelength ranges.  This measurement is the most reliable way to compare the sodium absorption depths between the two transits taken with different grisms, as the same wavelength regions can be compared directly.  We use this measurement to assess the significance of the sodium feature.  We found significant and consistent absorption depths between the two transits with different grating settings, measuring absorption depths of $47 \pm 11 \times 10^{-5}$ for the R500B observation and $49 \pm 17 \times 10^{-5}$ for the R500R observation (see Figure \ref{Figure:Nafig}).  The combined relative absorption depth of the Na~I feature using the mean spectrum is $47 \pm 9 \times 10^{-5}$, which is consistent with the spectral results.

\section{DISCUSSION}
\subsection{Atmospheric model fits}

The spectrum appears flat around the sodium feature, suggesting that the doublet must be unresolved in the data, and narrower than 50 \AA. Such a narrow feature could be indicative of haze or clouds, which obscure the lower regions of the atmosphere by scattering signatures, as seen in HD~189733b  (\citealt{2008MNRAS.385..109P, 2008A&A...481L..83L, 2009ApJ...699..478D, 2011A&A...526A..12D, 2011MNRAS.416.1443S, 2012MNRAS.422.2477H}).  Alternatively, in clear atmospheres dominated by alkali metal absorption, flat spectral regions can be observed where the abundance drops suddenly as altitude increases which counteracts the increase in cross-section toward the line core.  This behaviour is seen in HD~209458b (\citealt{2008ApJ...686..667S, 
 2011A&A...527A.110V}).  Abundance drops could be attributed to processes such as ionisation, or condensation if the temperature profile crosses the condensation curve.  We fit the spectrum of XO-2b with different models to assess the significance of the observed lack of line wings.

\begin{figure}
\centering
\includegraphics[width=8cm]{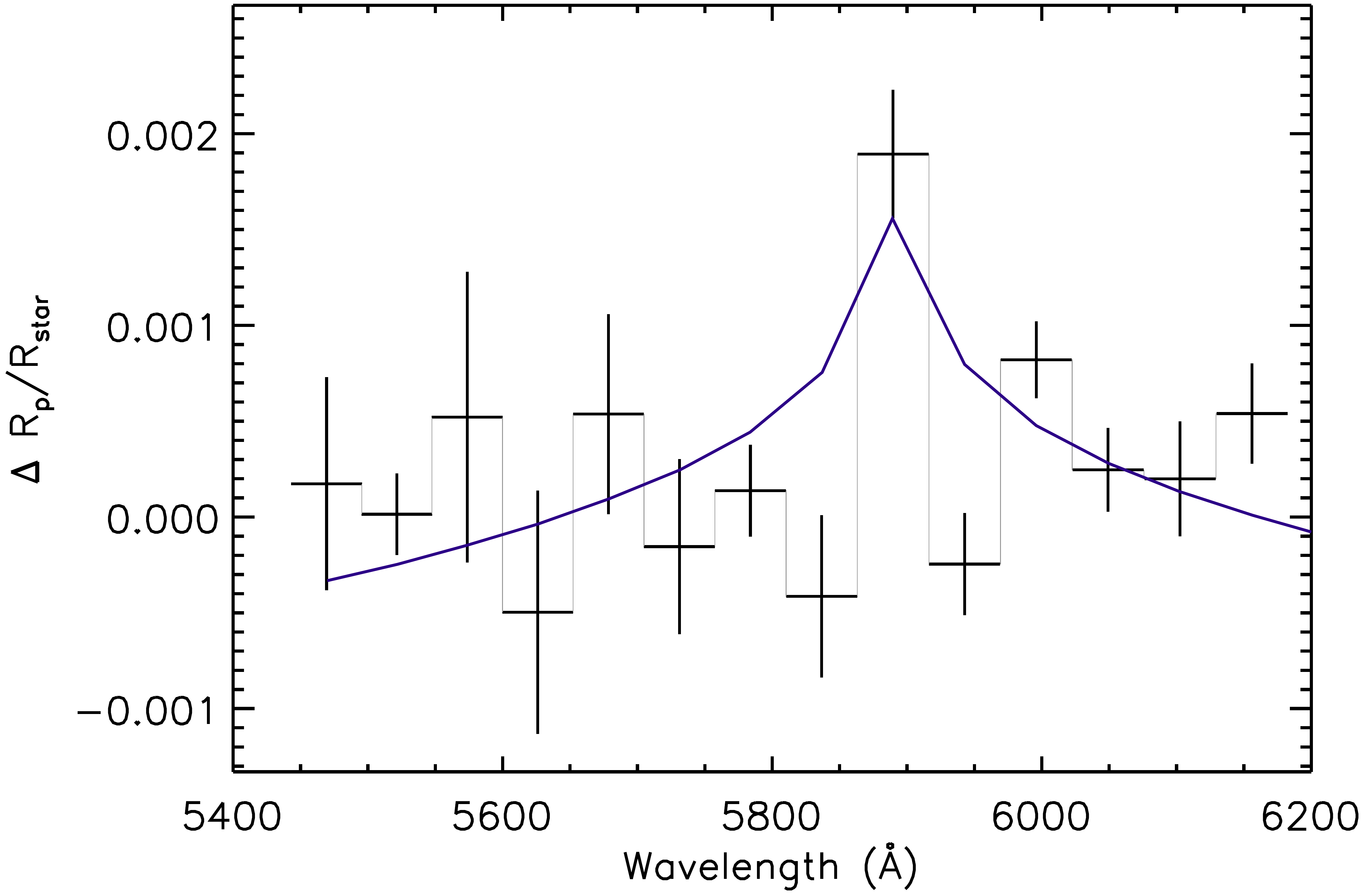}
\caption{The differential transmission spectrum from the R500R and R500B spectra around the sodium feature (black circles).  Also plotted (blue solid line) is a model fit from \citet{2010ApJ...709.1396F}, assuming a haze/cloud free atmosphere dominated in the optical by alkali metal absorption.}
\label{fig_notflat_fortney_fits}
\end{figure}

We first used a model transmission spectrum developed specifically for XO-2b from \cite{2011A&A...527A..73S} which is calculated using the 
hot-Jupiter model atmospheric calculations of \cite{
2010ApJ...709.1396F} and \cite{2011ApJ...727...65S}. 
These models include a self-consistent treatment of radiative transfer and
chemical equilibrium of neutral species, however the formation of hazes and photo-ionisation are both neglected.   
Broad sodium and potassium line wings are seen, each over 1000 \AA $ $ wide. The model was binned to the transmission spectral resolution of 50 \AA\ and fitted to the data with the reference level $z_0$ as a free parameter, as shown in Figure \ref{fig_notflat_fortney_fits}. The best fit gives $\chi^2=37.6$ for 13 DOF. 

We then modelled the feature using an analytic isothermal model transmission spectrum for the sodium feature as described in \citet{2012MNRAS.422.2477H}, based on the formalism of \citet{2008A&A...481L..83L} and \citet{2011A&A...527A.110V}, with temperature and the reference level, $z_0$ as free parameters. Fixing the isothermal model temperature to an average temperature of 800~K produced a fit very similar to the full model, with $\chi^2=38.3$ for 13 DOF.  Adding pressure broadening to this model produced no noticeable difference in the fits. Allowing the temperature parameter to vary freely produced a temperature of $390\pm120$~K ($\chi^2=31.6$ for 12 DOF) which is clearly too cold, showing that the model has to flatten the line wings to fit the data.  Neither model provides a satisfactory fit.

\begin{figure}
\centering
\includegraphics[width=8cm]{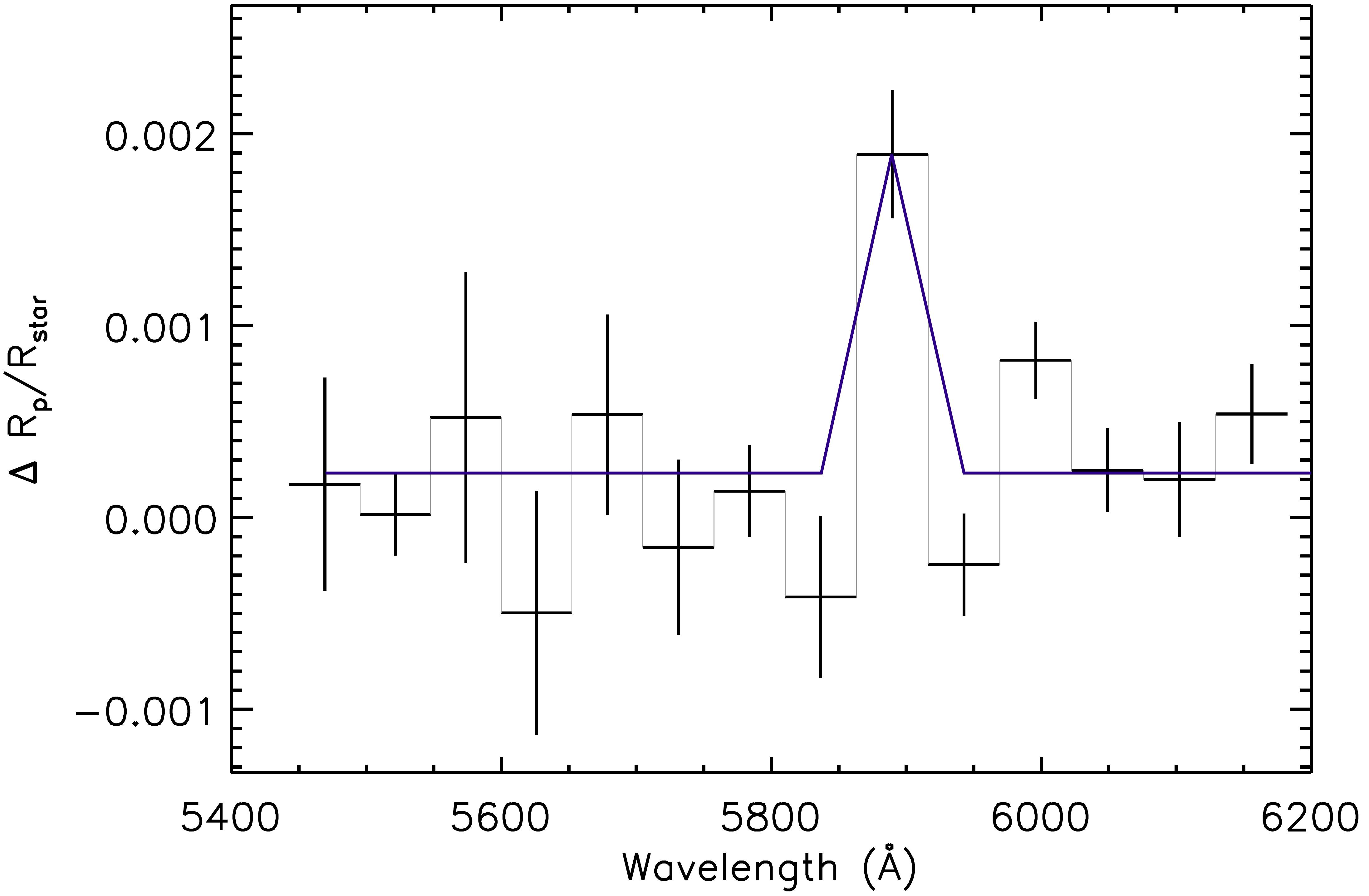}
\caption{Fit assuming an unresolved feature narrower than 50 \AA $ $ wide, either due to a cloud or haze layer obscuring the line wings, an abundance drop with increasing altitude. Fitted data points are shown in black with circles and joined with dotted lines, and the model is shown with a blue solid line.}
\label{fig_flat_fits}
\end{figure}

For an alternative model, we assumed that the sodium feature is unresolved at a resolution of 50 \AA\ with the line wings either obscured by scattering from clouds or hazes, or flat due to a sudden abundance change.  We choose a width of 12~\AA\ for the unresolved feature, which must be less than 50~\AA, and then artificially set the regions of the high-resolution model outside this wavelength range to be flat with a constant radius. The flat regions of the model were set to the zero reference level and fit with a free parameter.  In this model, all regions outside the 12 \AA\ feature have the same radius, which is allowed to move up and down relative to the sodium feature. This is equivalent to adjusting the height of the cloud deck/haze layer, or adjusting the sodium abundance at the lower altitudes probed by the sodium line wings. Additionally, the $z_0$ parameter is still free, allowing the model spectral profile to freely move up and down as a whole. The resulting fits are the same when this procedure is employed on both the analytic isothermal models from \citet{2012MNRAS.422.2477H} and the numerical \cite{
2010ApJ...709.1396F} 
models, as the narrow sodium feature is only detected in one 50 \AA\ bin, providing no temperature information.  The fit is shown in Figure \ref{fig_flat_fits}.  The best fit gives a $\chi^2=19.2$ for 12 DOF, which is a better fit than the standard model at the $3.5-4 \sigma$ level. 

Changing the width of the unresolved feature, wider or narrower than 12 \AA\, did not change the best fit, since the free parameter for the altitude of the line wings adjusted to compensate. Additionally, we tried adding a Mie or Rayleigh scattering cross-section to the sodium line cross-section in the isothermal models, to try to distinguish between the two cases of an obscuring haze or a sudden abundance decrease from photo-ionisation or condensation.  We find that we cannot differentiate with any significance between these cases, with the resulting $\chi^2$ values only $1 \sigma$ different. 

Both condensation and ionisation have been inferred in hot Jupiter atmospheres (\citealt{2008ApJ...686..667S, 2011A&A...527A.110V}).  Although the global average theoretical terminator $T$-$P$ profile does not cross the condensation curve of Na into Na$_2$S, it is possible for temperatures to be low enough for this condensation on the planet's night side, thus causing a decrease in Na~I abundance in the upper atmosphere.

\section{CONCLUSION}

We detect absorption from the sodium doublet in the atmosphere of XO-2b at the $5.2 \sigma$ level using multi-object long-slit spectroscopy. This makes XO-2b the first extrasolar planet where both potassium and sodium have been detected in its atmosphere, features which have both been long predicted to be present in hot Jupiters.  The detection helps demonstrate the feasibility of using ground-based multi-object spectroscopy for detecting atmospheric features in transiting extrasolar planets.  While our analysis was limited to measuring differential transmission spectra, due to slit losses, the precisions achieved on absolute transit depths (200 ppm) would already be sufficient to detect features on nearly a dozen exoplanets with larger scale heights, such as Wasp-17b, and adopting wider slits should substantially increase the photometric precision.

Of the well studied planets so far, an emerging picture is far from simple, with observations of clouds or hazes and large abundance changes throughout the atmospheres detected in the wide class of hot-Jupiter exoplanets.  
Further observations of XO-2b will be able to confirm and help distinguish the cause of the detected lack of sodium line wings, as a full broad-band transmission spectra could pick up Rayleigh scattering from haze, for instance.  Additionally, these low-resolution observations would highly benefit from complementary high-resolution transmission spectroscopy around the sodium and potassium features, as the measured alkali metal absorption line profiles can place constraints on temperatures and abundances.

\section*{ACKNOWLEDGMENTS}
We thank the entire GTC staff and especially Antonio Cabrera-Lavers for his continued excellent work in executing this program.
D. K. Sing and C. M. Huitson both acknowledge support from STFC. 
This work was partially supported by the Spanish
Plan Nacional de Astronomía y Astrofísica under grant AYA2008-06311-C02-01. 

\footnotesize{
\bibliographystyle{mn2e} 
\bibliography{xo2,xo2_gtc} 
}
\end{document}